%% file: paper.tex
\documentclass[]{bytedance_seed}

\usepackage[toc,page,header]{appendix}

\usepackage{minitoc}
\usepackage{cleveref} 
\usepackage{subcaption}
\usepackage{booktabs}
\usepackage{tabularx}
\usepackage{makecell} 

\ifdefined\final
  \usepackage[disable]{todonotes}
\else
  \usepackage[textsize=tiny]{todonotes}
\fi
\input{macro}

\title{SeedProteo: Accurate De Novo All-Atom Design of Protein Binders}

\affiliation[1]{ByteDance Seed}
\contribution{Full author list in Contributions}
\abstract{
\input{sections/000abstract}
}

\date{Feb 24, 2026}

\checkdata[Correspondence]{Quanquan Gu (\url{quanquan.gu@bytedance.com})}
\checkdata[Project Page]{\url{https://seedfold.github.io/}}

\begin{document}

\maketitle

%\newpage
%\tableofcontents
%\newpage

\input{sections/010intro}

\input{sections/030method}

\input{sections/040experiments}

\input{sections/050invitro}
\input{sections/100conclusion}

\clearpage

\bibliographystyle{unsrt}
\bibliography{main}

\clearpage

\beginappendix

\input{sections/appendix}

\end{document}

%% file: macro.tex
\usepackage{natbib}
\usepackage{latexsym}

\usepackage{url}
\usepackage{amssymb}
\usepackage[utf8]{inputenc}
\usepackage{microtype}
\usepackage{booktabs}
\usepackage{pifont} 
\usepackage{multirow}
\usepackage{makecell}
\usepackage{paralist}
\usepackage{xspace}
\usepackage{color}
\usepackage{xcolor}
\usepackage{colortbl}
\usepackage{adjustbox}
\usepackage{hyperref} 
\usepackage[edges]{forest}
\usepackage{tikz} 
\usepackage{caption}
\usepackage{amsfonts}

\hypersetup{
    colorlinks,
    linkcolor={blue!80!black},
    citecolor={blue!80!black},
}
\tikzset{
    root/.style =             {align=center, text width=1cm, rounded corners=3pt, line width=0.3mm, fill=gray!10, draw=gray!80, font=\small},
    demographic/.style =         {align=center, text width=1.8cm, rounded corners=3pt, line width=0.3mm, fill=blue!10, draw=blue!80, font=\footnotesize},
    demographic_work/.style =    {align=center, text width=10cm, rounded corners=3pt, line width=0.3mm, fill=blue!10, draw=blue!0, font=\footnotesize},
    character/.style =         {align=center, text width=1.8cm, rounded corners=3pt, line width=0.3mm, fill=red!10, draw=red!80, font=\footnotesize},
    character_work/.style =    {align=center, text width=10cm, rounded corners=3pt, line width=0.3mm, fill=red!10, draw=red!0, font=\footnotesize},
    personalization/.style =           {align=center, text width=1.8cm, rounded corners=3pt, line width=0.3mm, fill=cyan!10, draw=cyan!80, font=\footnotesize},
    personalization_work/.style =      {align=center, text width=10cm, rounded corners=3pt, line width=0.3mm, fill=cyan!10, draw=cyan!0, font=\footnotesize},
    risk/.style =         {align=center, text width=1.8cm, rounded corners=3pt, line width=0.3mm, fill=orange!10, draw=orange!80, font=\footnotesize},
    risk_work/.style =    {align=center, text width=10cm, rounded corners=3pt, line width=0.3mm, fill=orange!10, draw=orange!0, font=\footnotesize},
}

\usepackage{CJK}

%% file: sections/000abstract.tex
We present SeedProteo, a diffusion-based model for \textit{de novo} all-atom protein design. We demonstrate how to repurpose a cutting-edge folding architecture into a powerful generative design framework by effectively integrating self-conditioning features. Extensive benchmarks highlight the model's capabilities across two distinct tasks: in unconditional generation, SeedProteo exhibits superior length generalization and structural diversity, maintaining robustness for long sequences and complex topologies; in binder design, it achieves state-of-the-art performance among open-source methods, attaining the highest in-silico design success rates, structural diversity and novelty. Finally, we validate SeedProteo through wet-lab assays on two therapeutic targets, achieving hit rates of 70\%--80\% and picomolar-level binding affinities, establishing leading results. To facilitate community adoption, we provide public access to SeedProteo via a webserver (\url{https://seedfold.io/proteinDesign}).

\vspace{2mm}
\begin{center}
    \includegraphics[width=0.9\linewidth]{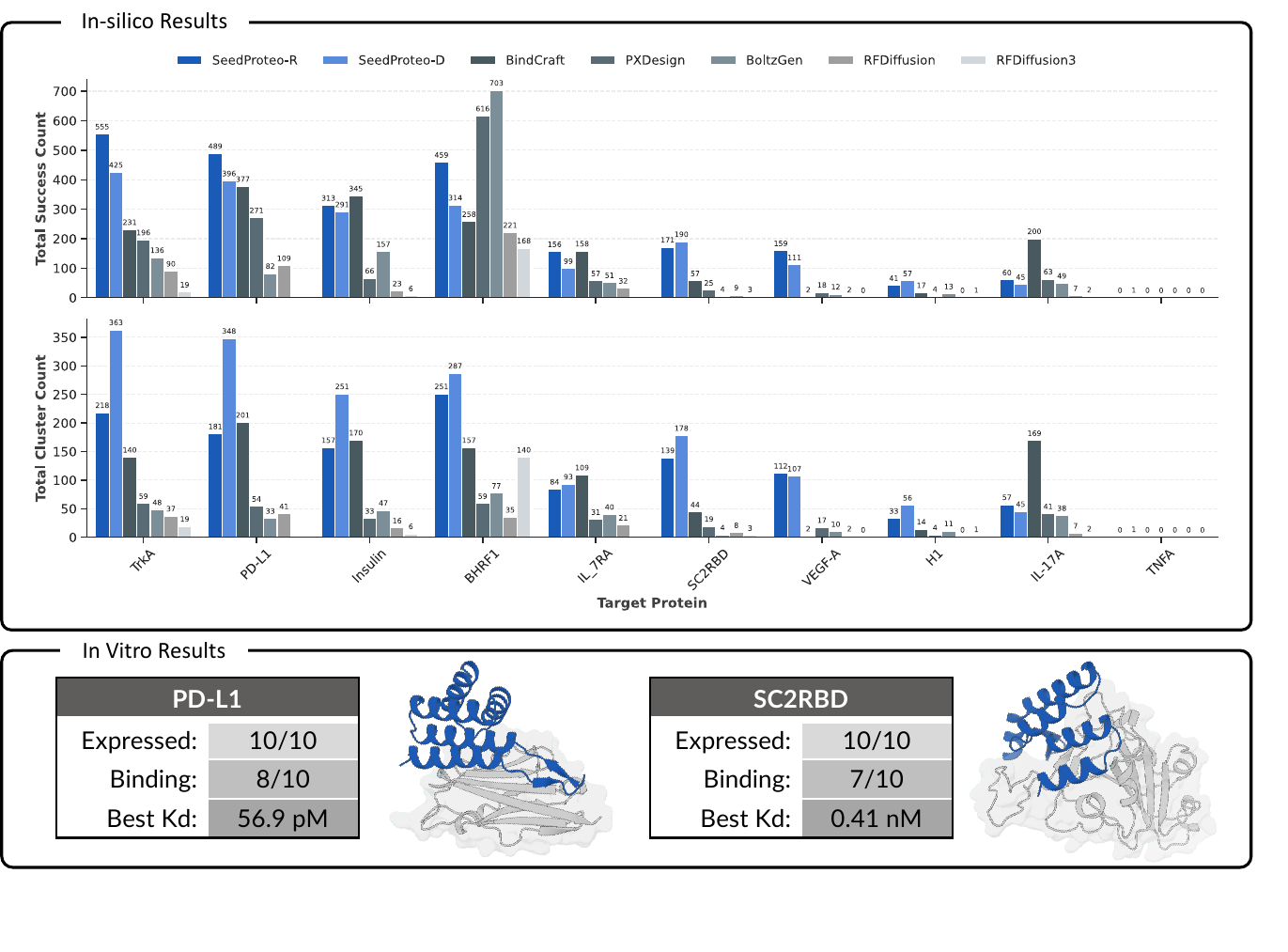}
    \label{fig:front}
\end{center}

\vspace{-10mm}

%% file: sections/010intro.tex
\vspace{-50pt}

\newpage
\section{Introduction}
Proteins perform a vast array of functions within organisms. As the evolution process spans millions to billions of years, the field of protein design aims to automate and accelerate it through computational methods. Since the function of proteins is confered via the structure, which is determined by the sequence, a line of research focuses on generating sequences \citep{dauparasRobustDeepLearning2022, zhengStructureinformedLanguageModels2023, alamdariProteinGenerationEvolutionary2024, wang2024dplm} and structures \cite{yimSE3DiffusionModel2023b, watsonNovoDesignProtein2023, ingrahamIlluminatingProteinSpace2023, zhang2025odesign}. Recently, a new paradigm has emerged emphasizing the co-generation of sequence and structure \cite{wang2025dplm2, hsieh2025dplm2_1, chen2025apm, multiflow}. However, improving sample diversity while preserving the coherence between modalities, specifically sequence-structure consistency, remains a formidable challenge. To address this, recent works have pivoted toward all-atom modeling \cite{quPallatomUnlockingNew2025, geffnerLaProteinaAtomisticProtein2025}. By directly designing proteins at the all-atom level, these methods naturally derive amino acid identities from atomic coordinates, achieving superior consistency. These methodological advancements have significantly propelled the field of \textit{de novo} protein design, particularly in binder design, the creation of mini-binders for specific therapeutic targets \cite{zambaldi2024novo}. This domain has witnessed a rapid evolution, progressing from classical hallucination-based methods \cite{pacesaOneshotDesignFunctional2025} or backbone-based design model \cite{zhang2025odesign, ren2025pxdesign} to the current wave of all-atom diffusion models \cite{butcher2025novo, chai2025zero, team2025latent}, all striving for enhanced designability and binding affinity.

In this study, we introduce \texttt{SeedProteo}, a diffusion-based model that learns the manifold of proteins at the all-atom level. SeedProteo mimics the architecture of AlphaFold3 \cite{abramsonAccurateStructurePrediction2024, wohlwendBoltz1DemocratizingBiomolecular2024, passaroAccurateEfficientBinding}, which has proven its effectiveness in protein folding. Through unifying the representation of different amino acids, we significantly enhance its generative power when the protein sequence and homology are not provided. A major issue in all-atom modeling is the backbone-sidechain inconsistency (i.e., the sequence-structure inconsistency), which means that the sidechain atoms are locally plausible but the derived sequence fails to fold into the global backbone structure. SeedProteo iteratively evolves the sequence and structure and integrates two components to enhance the modeling ability of two modalities:
(i) SeedProteo enhances the structural feasibility through the integration of geometric reasoning in the denoising process. To be specific, we reuse the denoised structure from the previous timestep \cite{chen2022analog} to stabilize the sampling process and conduct geometric reasoning in the pair representation space.
(ii) SeedProteo conducts energy minimization within the sequence space modeled by a Markov random field (MRF) \cite{metzler2005markov}. By integrating high-order couplings among distinct positions, we are able to sample globally feasible sequences and prevent being trapped in the local minima of simple patterns constituted by high-frequency amino acids.
We conduct extensive experiments on unconditional sampling, binder design, and in vitro validation. The results demonstrate that:

1) In unconditional generation, SeedProteo exhibits superior length generalization and structural diversity compared to state-of-the-art baselines. It maintains stable performance on long sequences (up to 1000 residues) and complex topologies, effectively addressing the length-dependent degradation issues observed in baseline diffusion models.

2) In binder design, to address the trade-off between design success rate and structural diversity, we introduce two inference modes: \texttt{SeedProteo-R} (Robust), which prioritizes longer secondary structure segments for higher design success, and \texttt{SeedProteo-D} (Diverse), which fosters topological richness. Experiments show robust performance across diverse targets, particularly excelling in challenging cases.

3) Furthermore, our analysis offers specific insights into how to achieve superior sequence-structure consistency within the \texttt{atom14} representation schema \cite{quPallatomUnlockingNew2025, butcher2025novo, starkUniversalBinderDesign}. We demonstrate that introducing a global MRF-based decoding module leads to more self-consistent outputs.

4) Beyond in-silico evaluation, we further validate SeedProteo with wet-lab experiments, demonstrating that designed binders exhibit strong binding in SPR assays with high hit rates across multiple therapeutic targets.

%% file: sections/030method.tex
\section{Method}
\subsection{Model Architecture}
SeedProteo is a diffusion-based all-atom protein generation model specifically designed for binder design. To enable direct modeling of all-atom structures, the primary challenge lies in representing side-chain atoms when amino acid types are unknown. To address this issue, SeedProteo introduces virtual atoms and adopts the \texttt{atom14} schema \cite{quPallatomUnlockingNew2025} for all amino acids, which includes four backbone atoms and ten side-chain atoms, with virtual atoms overlaid on the C$\alpha$ atom.

\begin{figure}[t]
    \centering
    \includegraphics[width=0.75\linewidth]{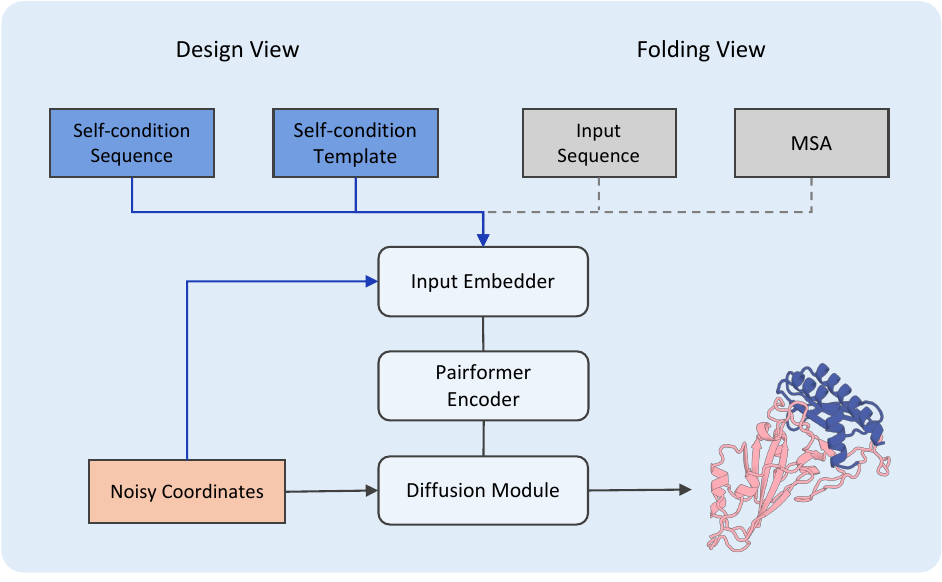}
    \caption{Overview of the SeedProteo framework. The right panel illustrates the input representation from a folding perspective. By modifying specific input channels within a nearly identical network architecture, we adapt the framework into the generative design model shown on the left.}
    \label{fig:model}
\end{figure}

As shown in Figure \ref{fig:model}, SeedProteo leverages the powerful non-equivariant atomic coordinate modeling capability of AF3-like folding model architectures, and therefore adopts a network structure consistent with such folding models: (i) an embedder module for feature initialization, (ii) an encoder composed of Pairformer blocks, and (iii) a diffusion module that denoises coordinates based on the encoder's representations.

However, a key challenge arises because the folding model's encoder is sequence-based. A trivial solution would be to replace the sequence with all \texttt{[MASK]} tokens, but such a degenerate biased input would render the encoder ineffective. To address this issue, SeedProteo additionally feeds noisy coordinates into the encoder to extract intrinsic geometric relationships: within the embedder, noisy coordinates are transformed into 1D sequence representation. This modification, however, introduces a computational cost—since the encoder, which contains cubic algorithmic complexity, becomes dependent on the noisy inputs. To mitigate the computational burden, we reduced the number of Pairformer layers from 48 to 12.

\subsection{Self-conditioning features}
The distribution of protein atomic coordinates constitutes a specialized manifold. Although diffusion models can directly learn this coordinate distribution, a simple diffusion process lacking constraints often leads to the generation of structures that are geometrically plausible yet physically unrealizable. These generated backbones often exhibit poor designability, meaning that no amino acid sequence can stably fold into the generated conformation. Consequently, sequences derived from inverse folding on these structures typically fail to adopt the intended topology. Recent advances in protein design methodologies have validated the effectiveness of self-conditioning. We posit that employing effective self-conditioning features represents a promising pathway for substantially enhancing design capabilities. The following features are utilized as conditioning inputs in our model.

\begin{enumerate}
    \item \textbf{Sequence Sampling:} SeedProteo employs a Markov Random Field (MRF)-based Sequence Module \cite{renAccurateRobustProtein2024, meynardfast} to decode amino acid sequences. The probability of a sequence $\mathbf{x}$ is parameterized using the learned pair and single representations (Eq. \ref{eq:mrf}):
    \begin{equation}
    \label{eq:mrf}
        \mathrm{P}(\mathbf{X} = \mathbf{x} \mid \mathbf{a}, \mathbf{z}) 
        \propto \exp \bigg[ \sum_{i=1}^{L} h_i(x_i \mid \mathbf{a}_i) + \sum_{i=1}^{L}\sum_{j=i+1}^{L} e_{ij} (x_i, x_j \mid \mathbf{z}_{ij}) \bigg],
    \end{equation} 
    where $\mathbf{x} = (x_1, \dots, x_L)$ denotes the target amino acid sequence of length $L$, and $\mathbf{a}, \mathbf{z}$ represent the learned single and pair representations, respectively. The term $h_i(\cdot)$ models the site-specific bias for amino acid $x_i$ conditioned on the single feature $\mathbf{a}_i$, while $e_{ij}(\cdot)$ captures the pairwise coupling propensity between residues $x_i$ and $x_j$ derived from the pair feature $\mathbf{z}_{ij}$ \cite{balakrishnanLearningGenerativeModels2011}.
    We observe that this sequence-guided approach enhances the model's co-design capability by mitigating potential conflicts between sequence and structural constraints. By integrating the MRF module, SeedProteo eliminates the need to infer amino acid types by translating \texttt{atom14} coordinates. Instead, it directly utilizes the sequence inferred by the MRF, which incorporates semantic information to avoid misclassification of structurally similar (isosteric) amino acids, as the final output sequence.

    \item \textbf{Secondary Structure Sequence:} SeedProteo accepts a secondary structure (SS) sequence \cite{hekkelman2025dssp} composed of three canonical characters (H, E, L) and a special mask token (X). The model performs unmasking predictions based on the input masked SS sequence and subsequently utilizes the predicted SS sequence for self-conditioning (Eq. \ref{eq:ss_seq}):
\begin{equation}
\label{eq:ss_seq}
\begin{aligned}
    \mathcal{S}^{(0)} &= \mathcal{S}_{\text{input}}  \\
    \mathcal{S}^{(t)}_{\text{pred}} &= f_{\theta}\left(\mathcal{S}^{(t-1)}_{\text{cond}}\right)\\
    \mathcal{S}^{(t)}_{\text{cond}} &= \alpha_t \cdot \mathcal{S}^{(t)}_{\text{pred}} + (1 - \alpha_t) \cdot \mathcal{S}^{(t-1)}_{\text{cond}},
\end{aligned}
\end{equation}
% \begin{equation}
% \label{eq:ss_seq}
% \begin{aligned}
%     \mathcal{S}^{(0)} &= \mathcal{S}_{\text{input}} \quad \text{(e.g., "XXXHHXXLXEEEE")} \\
%     \mathcal{S}^{(t)}_{\text{pred}} &= f_{\theta}\left(\mathcal{S}^{(t-1)}_{\text{cond}}\right) \quad \text{(e.g., "HHHHHLLLEEEEE")}\\
%     \mathcal{S}^{(t)}_{\text{cond}} &= \alpha_t \cdot \mathcal{S}^{(t)}_{\text{pred}} + (1 - \alpha_t) \cdot \mathcal{S}^{(t-1)}_{\text{cond}}
% \end{aligned}
% \end{equation}
where $\mathcal{S}_{\text{input}}$ represents the initial, potentially masked secondary structure representation. At iteration $t$, $f_{\theta}(\cdot)$ denotes the neural network predictor parameterized by $\theta$, yielding the unmasked prediction $\mathcal{S}^{(t)}_{\text{pred}}$. This prediction is integrated into the conditioning state $\mathcal{S}^{(t)}_{\text{cond}}$ via a scalar gating factor $\alpha_t \in [0, 1]$, which controls the update rate of the self-conditioning signal.
This capability enables SeedProteo to process arbitrarily generated SS sequences, even those containing unknown segments, and adaptively refine them into a physically plausible SS sequence. We demonstrate that incorporating secondary structure features effectively guides the generation of \textit{de novo}-like structures, significantly enhancing the success rate in binder design applications.

    \item \textbf{Structural Template:} SeedProteo utilizes the denoised structure obtained from the previous step as a structural template for self-conditioning. Specifically, a binning operation is applied to convert the C$_\beta$ distance map into a one-hot tensor, which serves as a pairwise feature. We demonstrate that incorporating such template features helps stabilize the structural sampling trajectory.

\end{enumerate}

\subsection{Training Objective}
SeedProteo employs a set of loss functions commonly used in AF3-like folding models \cite{abramsonAccurateStructurePrediction2024}, including a coordinate diffusion loss, a smooth LDDT loss, and a distogram loss. The coordinate diffusion loss is applied without pre-aligning the target and predicted structures, thereby compelling the model to learn equivariance directly through this loss. Additionally, a cross-entropy loss is used to supervise the MRF-sampled sequence and the predicted secondary structure sequence.

\subsection{Training Pipeline}
SeedProteo was trained from scratch. Inspired by the training strategies of folding models, we adopted a multi-stage training pipeline. In the first stage, the model was trained using a small crop size and strictly filtered monomer data. In the second stage, training continued with a larger crop size and an expanded dataset comprising comprehensively filtered monomer data. Meanwhile, motifs are incorporated during the training process with monomer data, where partial structures paired with corresponding sequences are provided to enhance the model’s responsiveness to conditional inputs. In the third stage, the large crop size was maintained, and training proceeded with a balanced 1:1 sampling ratio of monomer data and protein multimer data. Detailed descriptions of the datasets and hyperparameters are provided in (Appendix \ref{app:data_prep} and Table \ref{tab:training_stages}).

%% file: sections/040experiments.tex
\section{In-silico Results}
We evaluated SeedProteo on two primary tasks: unconditional protein monomer generation and protein binder design. Unconditional generation refers to designing monomeric proteins starting from Gaussian noise, which serves to assess the model's fundamental generative capability. Binder design, on the other hand, involves generating potential high-affinity binders based on the full atomic information of a given target protein, with the goal of enabling real-world applications.

Although numerous diffusion-based co-design methods are capable of jointly generating sequences and structures (in either backbone or all-atom form), existing results consistently indicate that the sequence consistency of their outputs is generally lower than that achieved by redesigning structures using ProteinMPNN \cite{dauparasRobustDeepLearning2022}. Therefore, in the subsequent evaluations, for the structures generated by diffusion-based methods, we employed ProteinMPNN for sequence redesign to ensure fair and comparable assessments. Regarding the details of the evaluation procedure and the definitions of the metrics, please refer to Appendix \ref{app:eval_settings}.

\begin{figure}[t]
    \centering
    \includegraphics[width=0.98\linewidth]{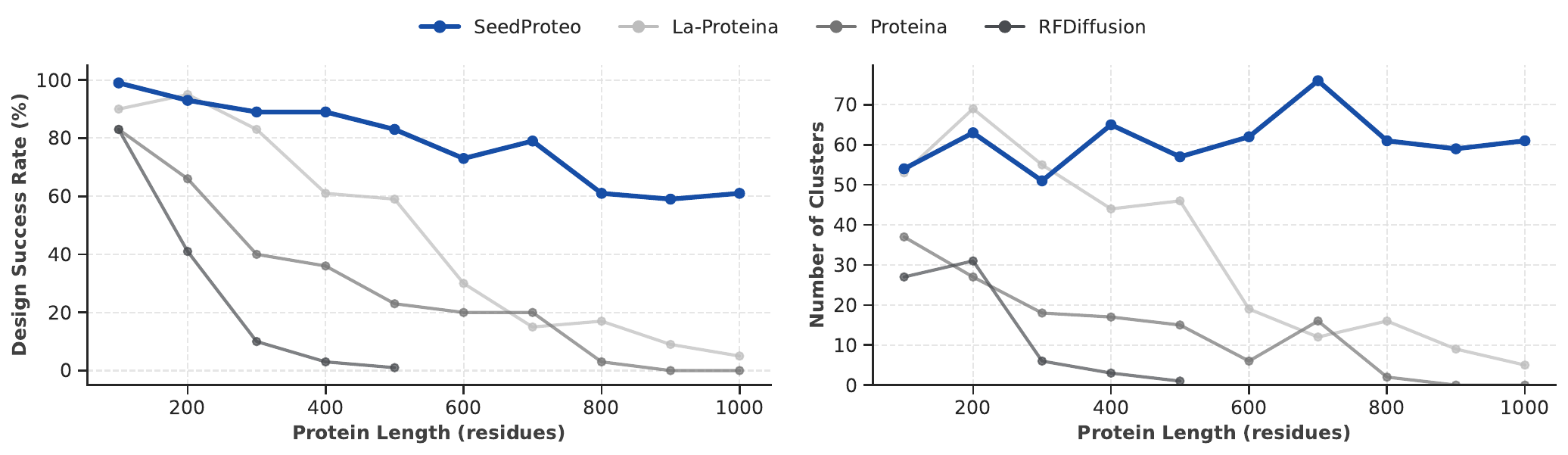}
    \caption{Unconditional monomer benchmark. Stricter thresholds are applied to define designability (in Appendix \ref{app:eval_settings}).}
    \label{fig:uncond_perf}
\end{figure}

\subsection{Unconditional Generation Benchmark}
\subsubsection{Scalability and Structural Diversity}
We first evaluate SeedProteo against state-of-the-art diffusion-based baselines, La-Proteina \cite{geffnerLaProteinaAtomisticProtein2025}, Proteina \cite{geffnerProteinaScalingFlowbased2024}, and RFDiffusion \cite{watsonNovoDesignProtein2023}, on the task of unconditional monomer generation. We focus on two critical dimensions: \textit{scalability} to long sequences and \textit{structural diversity}. As illustrated in Figure \ref{fig:uncond_perf} (Left), SeedProteo consistently achieves superior design success rates across the entire length spectrum (100--1000 residues). While competing methods exhibit a sharp performance degradation as sequence length increases, dropping to near-zero success rates beyond 600 residues, SeedProteo maintains robust performance, achieving a success rate of $>60\%$ even at length 1000. This suggests that our method effectively models the long-range dependencies and global constraints inherent in large protein assemblies, a capability that current baselines lack. In addition to quality, SeedProteo excels in generative diversity. Figure \ref{fig:uncond_perf} (Right) plots the number of unique structural clusters found within the successful samples. Notably, as protein length increases, baselines suffer from severe mode collapse (e.g., RFDiffusion and Proteina generate few to no unique clusters at long lengths). In contrast, SeedProteo sustains a high volume of unique clusters, indicating that it captures a richer and more generalizable distribution of the protein structure space.

\input{tables/uncond}

\input{tables/novelty}

\subsubsection{Fine-grained Evaluation across Structural Topologies}
Previous generative models often exhibit a strong preference for simple $\alpha$-helical architectures \cite{watsonNovoDesignProtein2023, quPallatomUnlockingNew2025}, frequently failing to generalize to complex topologies such as all-$\beta$ sheets. Our empirical observations confirm that designability is highly topology-dependent: primarily $\beta$-sheet structures consistently yield lower success rates compared to helical bundles across all models. To dissect this phenomenon, we present a fine-grained evaluation in Table \ref{tab:full_comparison}, breaking down performance by three distinct topologies: primarily $\alpha$-helix (HHH), primarily $\beta$-sheet (EEE) and hybrid Helix-Sheet-Loop (HEL). The results reveal a distinct advantage for SeedProteo in handling complex folds:

\begin{itemize}
    \item \textbf{Complex Topology Robustness:} In the Hybrid (HEL) category, SeedProteo achieves a 57\% success rate at length 1000, whereas baselines largely fail to produce valid structures beyond length 600.
    
    \item \textbf{$\beta$-Sheet Generation:} In the $\beta$-sheet (EEE) regime, which is notoriously difficult for generative models, SeedProteo is the only method capable of generating diverse, valid and novel structures (Success $>50\%$ and Novelty $<0.8$) at lengths up to 500. Other models suffer from complete generation failure or mode collapse in this category.
    
    \item \textbf{Novelty and Diversity:} Across all topologies, SeedProteo consistently achieves higher structural diversity (Unique Clusters) and favorable novelty scores, indicating that it does not merely retrieve training templates but effectively explores novel regions of the protein manifold.
\end{itemize}

\subsection{Binder Design Benchmark}
We benchmark SeedProteo against five leading open-source binder design methods: RFDiffusion \cite{watsonNovoDesignProtein2023}, RFDiffusion3 \cite{butcher2025novo}, BoltzGen \cite{starkUniversalBinderDesign}, PXDesign \cite{ren2025pxdesign}, and BindCraft \cite{pacesaOneshotDesignFunctional2025}. Following the protocol established in AlphaProteo \cite{zambaldi2024novo}, we generate approximately 1,000 candidates for each of the 10 benchmark targets, uniformly covering the target-specific binder length ranges. For structure-generating diffusion models (SeedProteo, RFDiffusion variants, PXDesign, and BoltzGen), binder sequences are redesigned using ProteinMPNN. A notable exception is BindCraft, a hallucination-based approach \cite{anishchenkoNovoProteinDesign2021}, which optimizes sequences directly via gradient descent on folding metrics (e.g., ipTM). All designs are subsequently evaluated \textit{in silico} using SeedFold \cite{seedfold}. We adopt the success criteria defined in AlphaProteo: minimum inter-chain Predicted Aligned Error (PAE) $\le 1.5$, binder pTM $\ge0.8$, and complex RMSD $<2.5$\AA.

In SeedProteo, we introduce two distinct Secondary Structure (SS) sampling modes to balance the trade-off between designability and structural diversity: 
\begin{itemize} 
    \item \textbf{SeedProteo-R:} This is a SS robust mode. Prioritizes fewer but longer continuous secondary structure segments. This mode imposes stronger structural stability constraints, typically yielding a higher design success number. 
    
    \item \textbf{SeedProteo-D:} This is a SS diverse mode. Allows for a higher frequency of shorter secondary structure segments. This mode encourages more complex topologies, resulting in greater structural diversity at the cost of slightly reduced raw success number. 
\end{itemize}

\paragraph{Performance Analysis.} We evaluated the unconditional generation capabilities of SeedProteo across a diverse set of ten target proteins. As shown in Figure \ref{fig:front}, SeedProteo-R demonstrates a commanding lead in total success counts across the benchmark. Notably, this advantage is most pronounced on challenging targets, specifically SC2RBD, VEGF-A, H1, and TNF$\alpha$.

\begin{figure}[t]
    \centering
    \includegraphics[width=0.90\linewidth]{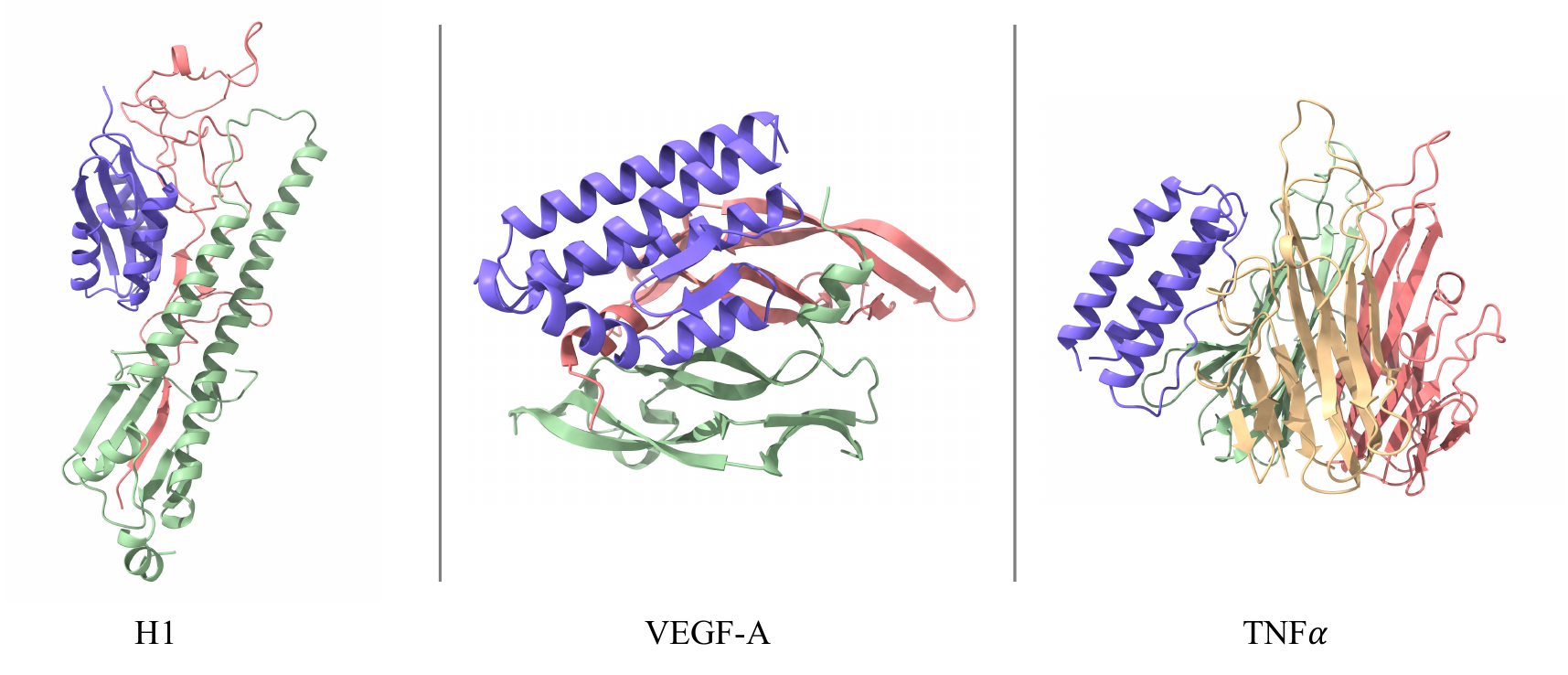}
    \caption{Binders generated by SeedProteo for the challenging multi-chain targets H1 (dimer), VEGF-A (dimer), and TNF-$\alpha$ (trimer). All displayed binders, colored in purple, meet the in silico success criteria.}
    \label{fig:case}
\end{figure}

Beyond design success, structural diversity is a critical metric for downstream applicability. Figure \ref{fig:front} highlights that \texttt{SeedProteo-D} excels in generating structurally distinct binders, achieving the highest number of unique success clusters in 8 out of the 10 targets. Complementing this structural diversity, Table \ref{tab:novelty} demonstrates that SeedProteo-D also achieves the best novelty scores for successfully designed binders in 8 out of the 10 targets. Notably, it consistently outperforms all baselines, except RFDiffusion3, across every target; for example, while baselines like PXDesign often yield scores above 0.9, SeedProteo-D maintains superior novelty with scores in the 0.8 range. This indicates that SeedProteo does not merely converge on a single solution but effectively explores the structural landscape to provide a diverse portfolio of high-quality and novel candidates. Figure \ref{fig:case} presents case studies on challenging targets.

\input{tables/MRF_binder}

\input{tables/ablation}

\subsection{Discussion on Co-Design Strategies}
We here discuss the significance of co-design and methods for enhancing co-design capabilities. While two-stage pipelines (backbone generation followed by inverse folding) have dominated recent protein design benchmarks, our analysis suggests that simultaneous sequence-structure generation (co-design) offers potential advantages for complex targets.

\paragraph{Beyond the Inverse Folding Bottleneck}
Empirical results (Figure \ref{fig:front} and Table \ref{tab:codesign_benchmark}) indicate that, on average, sequences derived from ProteinMPNN achieve higher success rates than those generated via pure co-design. This is expected, as ProteinMPNN is highly optimized for fixed-backbone recovery. However, we argue that reliance on inverse folding alone is not the ultimate solution for de novo binder design.

A striking exception is observed with the TNF$\alpha$ target. Here, ProteinMPNN-based methods almost universally fail (0 or 1 success) shown in Figure ~\ref{fig:front}, whereas our co-design approach yields valid binders. We reason that inverse folding models tend to fall into "safe" local minima, generating sequences with repetitive motifs (e.g., excessive electrostatic patterns like poly-EK or hydrophobic poly-A stretches) that optimize theoretical likelihood but fail restricted biophysical binding constraints for difficult targets. In contrast, co-design learns to exploit more complex sequence patterns that are still compatible with the structural constraints of the backbone, allowing the model to navigate out of the structural bottlenecks that trap fixed-backbone sequence designers.

\paragraph{Decoding Mechanisms}
We further compare the co-design mechanisms of SeedProteo with diffusion-based baselines, including RFDiffusion3 and BoltzGen, while excluding PXDesign and RFDiffusion as they are limited to backbone design rather than all-atom modeling. Both RFDiffusion3 and BoltzGen utilize variants of the \texttt{atom14} schema, relying on removing specific virtual atoms to implicitly decode amino acid types. In contrast, SeedProteo employs \texttt{atom14} strictly as a geometric input representation, while the final sequence identity is inferred via a Markov Random Field (MRF) module. This allows our model to capture higher-order residue dependencies that simple geometric decoding might miss.

\subsection{Ablation Study}
\paragraph{Impact of Secondary Structure Constraints}
To evaluate the specific contribution of secondary structure (SS) constraints, we compare our approach against a baseline termed \texttt{SeedProteo-M}, which performs binder design using fully masked SS sequences as condition. As shown in Table \ref{tab:codesign_benchmark}, incorporating secondary structure (SS) sequences as conditioning constraints significantly enhances design success. This constraint effectively prunes the search space, guiding the diffusion process toward \textit{de novo}-like fold topologies.

\paragraph{Impact of MRF module}
To validate the effectiveness of our sequence decoding module, we conducted a supplementary ablation study on unconditional generation (Table \ref{tab:ablation_mrf}). We established a baseline that predicts amino acid identities by aggregating features from the \texttt{atom14} representation at each residue position, a strategy adopted by the recent method \cite{quPallatomUnlockingNew2025}.

As shown in Table \ref{tab:ablation_mrf}, while the baseline achieves competitive performance on short sequences (Length 100), its capability degrades significantly as protein size increases. We reason this divergence to the inherent limitations of the decoding receptive fields. The \texttt{atom14} schema relies on recognizing amino acid types via local side-chain geometry (implicit conformational recognition). However, the underlying AF3-based architecture primarily models pairwise inductive biases and may lack the capacity to capture long-range atomic dependencies solely through local atoms aggregation. Consequently, the baseline struggles to maintain global consistency in larger systems. Conversely, the MRF module explicitly leverages global pairwise features during inference. By modeling the sequence distribution conditional on the entire backbone topology, the MRF ensures that the generated sequence is self-consistent with long-range structural constraints, thereby preventing the structural degeneration observed in the baseline at longer lengths.

%% file: tables/uncond.tex
\begin{table}[p]
    \centering
    \caption{\textbf{Unconditional Design Benchmark Performance across Structural Topologies.} 
    We compare \textbf{SeedProteo} against baselines on three different fold types: $\alpha$-helix bundles (HHH), Hybrid (HEL), and $\beta$-sheets (EEE). 
    Metrics reported in order: Design Success Rate (\textbf{Succ.}, $\uparrow$), Unique Success Clusters (\textbf{Uniq.}, $\uparrow$), and Novelty (\textbf{Nov.}, $\downarrow$). 
    The best performance is highlighted in \textbf{bold}. Dash (--) indicates model failure or insufficient successful samples.}
    \label{tab:full_comparison}
    
    \resizebox{\textwidth}{!}{%
    \begin{tabular}{l | ccc | ccc | ccc | ccc}
        \toprule
        \multicolumn{13}{c}{\large \textbf{Panel A: Primarily $\alpha$-Helix (HHH) Structures}} \\
        \midrule
        \multirow{2}{*}{\textbf{Length}} & \multicolumn{3}{c|}{\textbf{SeedProteo (Ours)}} & \multicolumn{3}{c|}{\textbf{La-Proteina}} & \multicolumn{3}{c|}{\textbf{Proteina}} & \multicolumn{3}{c}{\textbf{RFDiffusion}} \\
        \cmidrule(lr){2-4} \cmidrule(lr){5-7} \cmidrule(lr){8-10} \cmidrule(lr){11-13}
         & Succ. $\uparrow$ & Uniq. $\uparrow$ & Nov. $\downarrow$ & Succ. $\uparrow$ & Uniq. $\uparrow$ & Nov. $\downarrow$ & Succ. $\uparrow$ & Uniq. $\uparrow$ & Nov. $\downarrow$ & Succ. $\uparrow$ & Uniq. $\uparrow$ & Nov. $\downarrow$ \\
        \midrule
        100 & \textbf{100\%} & 17 & 0.89 & 93\% & \textbf{27} & 0.86 & 88\% & 22 & \textbf{0.83} & 87\% & 19 & 0.89 \\
        200 & \textbf{96\%} & 24 & 0.85 & \textbf{96\%} & \textbf{60} & \textbf{0.80} & 73\% & 21 & 0.83 & 73\% & 8 & 0.85 \\
        300 & \textbf{95\%} & 24 & 0.83 & 85\% & \textbf{47} & \textbf{0.80} & 51\% & 13 & 0.88 & -- & -- & -- \\
        400 & \textbf{98\%} & 32 & 0.79 & 61\% & \textbf{40} & \textbf{0.76} & 39\% & 14 & 0.88 & -- & -- & -- \\
        500 & \textbf{90\%} & 36 & 0.81 & 67\% & \textbf{43} & \textbf{0.74} & 33\% & 12 & 0.85 & -- & -- & -- \\
        600 & \textbf{86\%} & \textbf{38} & \textbf{0.79} & 43\% & 9 & 0.88 & 27\% & 6 & 0.89 & -- & -- & -- \\
        700 & \textbf{86\%} & \textbf{49} & \textbf{0.82} & 38\% & 6 & 0.83 & 24\% & 15 & 0.89 & -- & -- & -- \\
        800 & \textbf{65\%} & \textbf{39} & \textbf{0.79} & 41\% & 10 & 0.86 & 4\% & 2 & 0.96 & -- & -- & -- \\
        900 & \textbf{66\%} & \textbf{45} & \textbf{0.75} & 41\% & 7 & 0.88 & -- & -- & -- & -- & -- & -- \\
        1000 & \textbf{63\%} & \textbf{41} & \textbf{0.77} & 31\% & 5 & 0.88 & -- & -- & -- & -- & -- & -- \\
        
        \midrule
        \multicolumn{13}{c}{\large \textbf{Panel B: Hybrid Helix-E-Loop (HEL) Structures}} \\
        \midrule
        \multirow{2}{*}{\textbf{Length}} & \multicolumn{3}{c|}{\textbf{SeedProteo (Ours)}} & \multicolumn{3}{c|}{\textbf{La-Proteina}} & \multicolumn{3}{c|}{\textbf{Proteina}} & \multicolumn{3}{c}{\textbf{RFDiffusion}} \\
        \cmidrule(lr){2-4} \cmidrule(lr){5-7} \cmidrule(lr){8-10} \cmidrule(lr){11-13}
         & Succ. $\uparrow$ & Uniq. $\uparrow$ & Nov. $\downarrow$ & Succ. $\uparrow$ & Uniq. $\uparrow$ & Nov. $\downarrow$ & Succ. $\uparrow$ & Uniq. $\uparrow$ & Nov. $\downarrow$ & Succ. $\uparrow$ & Uniq. $\uparrow$ & Nov. $\downarrow$ \\
        \midrule
        100 & \textbf{97\%} & 26 & \textbf{0.82} & 87\% & \textbf{27} & 0.86 & 87\% & 13 & \textbf{0.82} & 71\% & 12 & 0.85 \\
        200 & \textbf{92\%} & \textbf{32} & \textbf{0.79} & 90\% & 7 & 0.81 & 55\% & 6 & 0.82 & 38\% & 22 & 0.80 \\
        300 & \textbf{81\%} & \textbf{21} & \textbf{0.77} & 72\% & 8 & 0.83 & 15\% & 4 & 0.85 & 11\% & 6 & 0.78 \\
        400 & \textbf{79\%} & \textbf{26} & \textbf{0.77} & 62\% & 3 & 0.79 & 12\% & 2 & 0.91 & 3\% & 3 & 0.80 \\
        500 & \textbf{69\%} & \textbf{20} & \textbf{0.75} & 60\% & 3 & 0.83 & 5\% & 2 & 0.87 & 1\% & 1 & 0.76 \\
        600 & \textbf{57\%} & \textbf{23} & \textbf{0.77} & 26\% & 12 & 0.85 & -- & -- & -- & -- & -- & -- \\
        700 & \textbf{72\%} & \textbf{28} & \textbf{0.77} & 9\% & 6 & 0.88 & 9\% & 1 & 0.93 & -- & -- & -- \\
        800 & \textbf{56\%} & \textbf{22} & \textbf{0.79} & 11\% & 6 & 0.86 & -- & -- & -- & -- & -- & -- \\
        900 & \textbf{44\%} & \textbf{14} & \textbf{0.78} & 3\% & 2 & 0.89 & -- & -- & -- & -- & -- & -- \\
        1000 & \textbf{57\%} & \textbf{20} & \textbf{0.74} & -- & -- & -- & -- & -- & -- & -- & -- & -- \\
        
        \midrule
        \multicolumn{13}{c}{\large \textbf{Panel C: Primarily $\beta$-Sheet (EEE) Structures}} \\
        \midrule
        \multirow{2}{*}{\textbf{Length}} & \multicolumn{3}{c|}{\textbf{SeedProteo (Ours)}} & \multicolumn{3}{c|}{\textbf{La-Proteina}} & \multicolumn{3}{c|}{\textbf{Proteina}} & \multicolumn{3}{c}{\textbf{RFDiffusion}} \\
        \cmidrule(lr){2-4} \cmidrule(lr){5-7} \cmidrule(lr){8-10} \cmidrule(lr){11-13}
         & Succ. $\uparrow$ & Uniq. $\uparrow$ & Nov. $\downarrow$ & Succ. $\uparrow$ & Uniq. $\uparrow$ & Nov. $\downarrow$ & Succ. $\uparrow$ & Uniq. $\uparrow$ & Nov. $\downarrow$ & Succ. $\uparrow$ & Uniq. $\uparrow$ & Nov. $\downarrow$ \\
        \midrule
        100 & \textbf{100\%} & \textbf{12} & \textbf{0.80} & \textbf{100\%} & 1 & 0.88 & 50\% & 3 & 0.84 & -- & -- & -- \\
        200 & 86\% & \textbf{6} & \textbf{0.80} & \textbf{100\%} & 2 & 0.84 & -- & -- & -- & 20\% & 1 & 0.75 \\
        300 & \textbf{80\%} & \textbf{4} & \textbf{0.80} & -- & -- & -- & 25\% & 1 & 0.85 & -- & -- & -- \\
        400 & \textbf{73\%} & \textbf{7} & \textbf{0.80} & 50\% & 1 & 0.85 & 57\% & 2 & 0.85 & -- & -- & -- \\
        500 & \textbf{50\%} & \textbf{1} & \textbf{0.68} & -- & -- & -- & 50\% & 1 & 0.74 & -- & -- & -- \\
        600 & -- & -- & -- & \textbf{25\%} & \textbf{4} & \textbf{0.86} & -- & -- & -- & -- & -- & -- \\
        700 & -- & -- & -- & \textbf{8\%} & \textbf{1} & \textbf{0.86} & -- & -- & -- & -- & -- & -- \\
        \bottomrule
    \end{tabular}%
    }
    \vspace{0.5em}
    \\
    \footnotesize{\textit{Note: Only lengths up to 700 are shown for EEE as all methods failed to generate valid long $\beta$-sheet structures.}}
\end{table}

%% file: tables/novelty.tex
\begin{table}[t]
\centering
\setlength{\tabcolsep}{4.5pt}
\renewcommand{\arraystretch}{1.0}

\caption{\textbf{Structural Novelty ($\downarrow$) in Binder Design Benchmark.} Transposed comparison of average novelty scores. We compare \textbf{SeedProteo} variants against baselines across ten targets. \textbf{Lower is Better.} Best performance per target is highlighted in \textbf{bold}, and the second best is \underline{underlined}.}
\label{tab:novelty}

\begin{tabular}{l cccccccccc}
\toprule
\textbf{Method} & \textbf{TrkA} & \textbf{PD-L1} & \textbf{Insulin} & \textbf{BHRF1} & \textbf{IL-7RA} & \textbf{SC2RBD} & \textbf{VEGF-A} & \textbf{H1} & \textbf{IL-17A} & \textbf{TNFA} \\
\midrule
\multicolumn{11}{l}{\textit{\textbf{Ours}}} \\
SeedProteo-D & \underline{0.829} & \textbf{0.832} & \textbf{0.837} & \textbf{0.822} & \textbf{0.840} & \textbf{0.819} & \textbf{0.836} & \textbf{0.823} & \underline{0.806} & \textbf{0.870} \\
SeedProteo-R & 0.905 & 0.913 & 0.911 & 0.872 & 0.917 & \underline{0.858} & 0.901 & 0.890 & 0.855 & - \\
\midrule
\multicolumn{11}{l}{\textit{\textbf{Baselines}}} \\
BindCraft    & 0.849 & 0.856 & \underline{0.864} & 0.847 & \underline{0.861} & 0.863 & \underline{0.850} & \underline{0.830} & 0.818 & -- \\
PXDesign     & 0.914 & 0.929 & 0.928 & 0.924 & 0.928 & 0.917 & 0.913 & 0.888 & 0.906 & -- \\
BoltzGen     & 0.908 & 0.924 & 0.929 & 0.928 & 0.885 & 0.915 & 0.902 & 0.885 & 0.863 & -- \\
RFDiffusion  & 0.932 & 0.934 & 0.927 & 0.946 & 0.916 & 0.912 & 0.940 & --    & 0.938 & -- \\
RFDiffusion3 & \textbf{0.808} & \underline{0.834} & 0.876 & \underline{0.845} & \textbf{0.840} & -- & -- & 0.930 & \textbf{0.800} & -- \\
\bottomrule
\end{tabular}
\end{table}

%% file: tables/MRF_binder.tex
\begin{table}[t]
\centering
\caption{\textbf{Benchmarking Co-design Capabilities of Diffusion-based Models for Binder Design.}}
\label{tab:codesign_benchmark}
\resizebox{\textwidth}{!}{%
\begin{tabular}{lcccccccccc}
\toprule
\textbf{Method} & \textbf{BHRF1} & \textbf{SC2RBD} & \textbf{IL-7RA} & \textbf{PD-L1} & \textbf{TrkA} & \textbf{Insulin} & \textbf{H1} & \textbf{VEGF-A} & \textbf{IL-17A} & \textbf{TNF$\alpha$} \\
\midrule
\multicolumn{11}{l}{\textit{Baselines}} \\
BoltzGen & \textbf{627} & 0 & 10 & 17 & 21 & 99 & 1 & 1 & 5 & 0 \\
RFDiffusion3 & 9 & 0 & 0 & 1 & 0 & 0 & 0 & 0 & 0 & 0 \\
\midrule
\multicolumn{11}{l}{\textit{SeedProteo (Ours)}} \\
SeedProteo-R & 296 & \textbf{92} & \textbf{100} & \textbf{380} & \textbf{232} & \textbf{303} & 16 & \textbf{127} & \textbf{47} & 1 \\
SeedProteo-D & 139 & 80 & 52 & 265 & 143 & 181 & \textbf{25} & 45 & 17 & \textbf{3} \\
SeedProteo-M & 133 & 77 & 25 & 154 & 143 & 189 & 12 & 47 & 13 & 1 \\
\bottomrule
\end{tabular}%
}
\end{table}

%% file: tables/ablation.tex
\begin{table}[t]
\centering
\caption{\textbf{Ablation Study on Sequence Decoding Strategies.}}
\label{tab:ablation_mrf}
\begin{tabular}{l l cccc}
\toprule
\textbf{Length} & \textbf{Decoding Method} & \textbf{scRMSD} $\downarrow$ & \textbf{scTM} $\uparrow$ & \textbf{pLDDT} $\uparrow$ & \textbf{Succ. Rate} $\uparrow$  \\
\midrule
\multirow{2}{*}{100} 
& Baseline (\texttt{atom14}) & \textbf{1.11} & \textbf{0.94} & 88.56 & \textbf{0.90} \\
& SeedProteo (MRF)  & 1.63 & 0.93 & \textbf{88.65} & 0.84 \\
\midrule
\multirow{2}{*}{200} 
& Baseline (\texttt{atom14}) & 3.87 & 0.86 & 78.06 & 0.68 \\
& SeedProteo (MRF)  & \textbf{2.08} & \textbf{0.92} & \textbf{84.36} & \textbf{0.80} \\
\midrule
\multirow{2}{*}{300} 
& Baseline (\texttt{atom14}) & 4.71 & 0.83 & 73.40 & 0.52 \\
& SeedProteo (MRF)  & \textbf{2.30} & \textbf{0.91} & \textbf{80.33} & \textbf{0.76} \\
\bottomrule
\end{tabular}%
\end{table}

%% file: sections/050invitro.tex
\section{In Vitro Experiments}

\begin{figure}[t]
    \centering
    \includegraphics[width=0.95\linewidth]{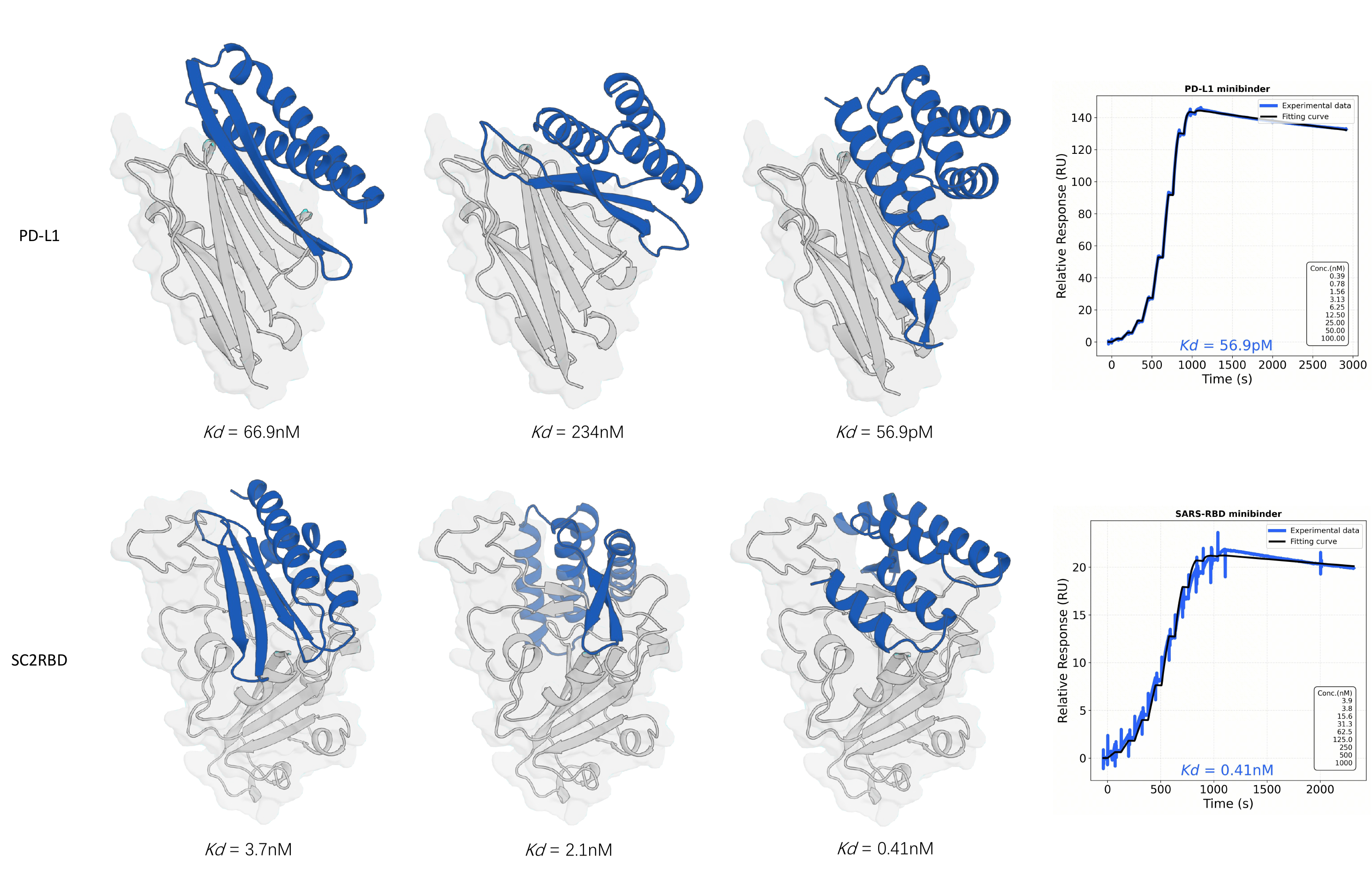}
    \caption{Wet-lab validation of SeedProteo-designed binders. For each target (PD-L1, top; SC2RBD, bottom), three representative binding complex structures are shown alongside the SPR sensorgram of the highest-affinity binder.}
    \label{fig:wetlab_res}
\end{figure}

To validate the practical utility of SeedProteo beyond in-silico metrics, we conducted wet-lab experiments on two therapeutically relevant protein targets: the SARS-CoV-2 receptor-binding domain (SC2RBD) \cite{cao2020novo} and Programmed Death-Ligand 1 (PD-L1) \cite{yang2025design}. These targets were selected due to their high clinical relevance---SC2RBD is a primary target for COVID-19 therapeutic antibody development, while PD-L1 is a critical immune checkpoint widely pursued in cancer immunotherapy---as well as their structural diversity, which allows us to assess the generalizability of our model across distinct binding interfaces.

For each target, we generated several thousand minibinders using SeedProteo. The structures passing the in-silico criteria were further clustered using Foldseek to ensure structural diversity, from which 10 representative minibinders were selected per target for experimental characterization. The selected minibinders were recombinantly expressed in \textit{E.~coli} with an N-terminal His-SUMO tag and purified via Ni-NTA affinity chromatography followed by size-exclusion chromatography (detailed protocols in Appendix~\ref{app:protein_production}). Binding properties were characterized using surface plasmon resonance (SPR) on a Biacore~8K+ instrument, where candidates were first screened at two concentrations and then subjected to kinetic affinity measurements using either single-cycle or multi-cycle protocols (Appendix~\ref{app:spr}). Protein expression and SPR assays for all designed candidates were performed by WuXi AppTec (China).

As shown in Figure~\ref{fig:wetlab_res}, SeedProteo achieved hit rates of 70\%--80\% across both targets, with the best binders reaching picomolar affinity for PD-L1 and sub-nanomolar affinity for SC2RBD. These results demonstrate that our in-silico design pipeline reliably produces binders with strong experimental binding, bridging the gap between computational prediction and wet-lab validation.

%% file: sections/100conclusion.tex
\section{Conclusion}

In this work, we presented SeedProteo, a diffusion-based all-atom protein design model. Our empirical benchmarks demonstrate that SeedProteo exhibits superior robustness across long sequences and complex topologies in unconditional design. Furthermore, it achieves state-of-the-art performance among open-source methods for binder design, particularly on hard and multi-chain targets. While effective co-design remains an open challenge in the field, we offer specific insights into achieving superior sequence-structure consistency via a global MRF module. Crucially, our wet-lab validation demonstrates that these in-silico advantages carry over to real-world efficacy, yielding experimentally confirmed binders with high hit rates and picomolar-level affinities. We hope these findings will inspire significant community efforts to further advance the frontier of simultaneous sequence-structure generation.

\newpage

\input{sections/revision}

\newpage

\section*{Contributions}

\textbf{Project Lead}

Wei Qu$^{1}$

% $^{*}$Equally Contributed

\textbf{Contributors}

Yiming Ma$^{1,2,\dagger}$, Fei Ye$^{1}$, Chan Lu$^{1}$, Yi Zhou$^{1}$, Kexin Zhang$^{1,3,\dagger}$, Lan Wang$^{1}$, Minrui Gui$^{1,4,\dagger}$

$^ \dagger$ Work is done during their internship at Bytedance Seed.

\textbf{Overall Technical Lead}

Quanquan Gu$^{1}$

\textbf{Affiliation}

$^1$~ByteDance Seed

$^2$~Peking University

$^3$~ShanghaiTech University

$^4$~University of California, Los Angeles

\section*{Acknowledgments}

We thank Liang Hong, Zaixiang Zheng, Xinyou Wang, Jiasheng Ye, Jing Yuan, Yilai Li, Zhenghua Wang, Yuning Shen, Cheng-Yen Hsieh, Huizhuo Yuan, as well as other colleagues at ByteDance for their discussions and support.

%% file: sections/revision.tex
\section{Revision History}
\subsection*{Version 2, February 24, 2026}

\begin{itemize} 
    \item Added in vitro experimental validation results.
    \begin{itemize}
        \item Added wet-lab experiments results with SPR binding characterization on SC2RBD and PD-L1 targets.
        \item Added revision history section.
    \end{itemize}
    \item Updated several citations.
\end{itemize}

\subsection*{Version 1, December 31, 2025}
Initial manuscript release.

%% file: sections/appendix.tex
\section{Data Preparation and Curation}
\label{app:data_prep}

Training a foundation model for protein design, particularly one capable of binder generation, necessitates a diverse corpus comprising both monomeric and multimeric structures. While predicted monomer databases such as the AlphaFold Database (AFDB) \cite{varadiAlphaFoldProteinStructure2022a} and ESMAtlas \cite{lin2023evolutionary} provide extensive coverage of single-chain structures, existing multimer datasets derived from the RCSB PDB \cite{berman2000protein} are limited in scale and often unsuitable for the specific two-chain requirements of binder design tasks. To address this, we constructed a composite training dataset by integrating curated monomer data with multimeric interactions sourced from Pinder \cite{kovtun2024pinder} and HumanPPI \cite{zhangPredictingProteinproteinInteractions2025}. We detail the processing and filtering pipelines for each category below.

\subsection{Monomer Data Curation}
Our monomeric dataset is derived exclusively from AFDB and ESMAtlas. For the ESMAtlas component, to mitigate potential quality issues associated with ESMFold \cite{lin2023evolutionary} predictions, we retrieved the corresponding sequences from the MGnify database \cite{richardson2023mgnify} and re-predicted their structures using AlphaFold2 (AF2) \cite{jumper2021highly}. 

To ensure high data quality, we applied the following inclusion criteria to all monomer structures:
\begin{itemize}
    \item Sequence length between 50 and 768 residues.
    \item Average pLDDT score $> 80$.
    \item Secondary structure coil fraction $< 50\%$.
\end{itemize}
To eliminate redundancy, we performed clustering using Foldseek \cite{vankempenFastAccurateProtein2024} and MMseqs2 \cite{steineggerMMseqs2EnablesSensitive2017} and retained only the cluster representatives (centroids). The final curated monomer dataset consists of approximately 0.5 million structures. Sequences that can be folded by SeedFold \cite{seedfold} in a single-sequence setting are exclusively allocated for the training of the initial phase.

\subsection{Multimer Data Curation}
Protein-Protein Interactions (PPIs) are fundamental to training a robust binder design model. We sourced multimer data from two primary streams to cover experimentally determined interactions and high-quality predicted domain interactions.

\paragraph{Experimental PPIs (Pinder).} 
To align our training data with the standard inference scenario, where a binder chain is designed for a specific target chain, we utilized the Pinder dataset. Using the curated holo-structure PDB IDs and chain IDs provided by Pinder, we extracted relevant substructures from RCSB biological assemblies. We filtered these two-chain complexes by requiring: (1) a coil fraction of $< 50\%$ for each chain, and (2) a minimum interfacial C$_\beta$ distance of $< 8\text{\AA}$ to ensure meaningful contact. Following structural clustering via Foldseek-Multimer, we retained approximately 50,000 unique cluster representatives.

\paragraph{Augmented DDI Dataset (HumanPPI).} 
To overcome the scarcity of experimental PPI data, we augmented our training set with domain-domain interactions (DDIs). This approach relies on the established hypothesis that intrachain DDI interfaces resemble interchain PPI interfaces in terms of coevolutionary and physicochemical properties \cite{lau2024exploring}. We utilized the HumanPPI database, which distills DDIs from accurately predicted monomer structures in the AFDB. We applied the identical filtering and clustering pipeline used for the Pinder dataset (coil check and distance constraints). This augmentation yielded an additional $\sim 0.1$ million interaction pairs.

\input{tables/training_pipeline}

\section{Evaluation Settings}
\label{app:eval_settings}

\subsection{Unconditional Generation Evaluation}
For unconditional monomer generation, we employ a self-consistency assessment pipeline to evaluate structural designability. For each generated backbone, we first only derive \textbf{one} amino acid sequence using ProteinMPNN (inverse folding). The sequence is subsequently refolded using SeedFold, an AlphaFold3-like folding model in a single-sequence setting. A design is considered successful if it satisfies two criteria: 
(1) the C$\alpha$-RMSD between the generated backbone and the refolded structure is $< 2.0$\AA, and 
(2) the average pLDDT score of the refolded structure is $> 80$. 
Metrics for \textbf{structural diversity} (number of unique structural clusters) and \textbf{novelty} (maximum TM-score relative to the PDB) are calculated with Foldseek \cite{vankempenFastAccurateProtein2024} exclusively on the subset of designable cases.

\paragraph{Topology-Aware Analysis.}
To address the long-standing challenge in protein generative models, specifically the bias toward $\alpha$-helical structures and the difficulty in generating valid $\beta$-sheets, we incorporate a fine-grained analysis based on secondary structure topology. Based on secondary structure proportions, we classify the generated proteins into three distinct fold types. We define \textbf{EEE} (primarily $\beta$-sheet) as structures with over 40\% sheet and under 20\% helix content, and \textbf{HHH} (primarily $\alpha$-helical) as those with over 50\% helix and under 10\% sheet content; both categories require a loop fraction below 45\%. All remaining structures are categorized as hybrid Helix-E-Loop (\textbf{HEL}).

\subsection{Binder Design Evaluation}
We evaluate binder design performance on a benchmark set of 10 targets, consistent with the validation set used in AlphaProteo (see Table \ref{tab:targets}). To rigorously assess the upper bound of design capability across different length scales, we adopt a dense sampling strategy: for each target, we sample 100 candidates at every 5-residue interval within the target-specific binder length range. For example, for the SC2RBD target (length range 80--120 residues), this results in a total of 900 sampled candidates.

For sequence recovery, we generate two sequences for each designed binder backbone using SolubleMPNN at a sampling temperature of $\tau = 0.001$. The designs are then validated using SeedFold in single sequence with target template mode. A binder design is considered successful if it meets the following criteria:
\begin{itemize}
    \item Interface Predicted Aligned Error (min PAE interaction) $< 1.5$;
    \item Binder pTM score $> 0.8$;
    \item Complex RMSD $< 2.5$\AA.
\end{itemize}

\input{tables/targets}

\section{Baseline Sampling}
\label{app:baseline_sampling}

In this section, we detail the specific configurations, checkpoints, and codebases used for all baseline methods included in our benchmarks.

\subsection{Unconditional Benchmark}

\textbf{La-Proteina.} We utilized the official implementation and checkpoints provided in the La-Proteina repository\footnote{\url{https://github.com/NVIDIA-Digital-Bio/la-proteina}}. To ensure optimal performance across different scales, we selected checkpoints based on the target sequence length: the \texttt{LD1\_ucond\_notri\_512.ckpt} checkpoint was used for lengths 100--500, while \texttt{LD3\_ucond\_notri\_800.ckpt} was employed for lengths 600--1000. All sampling was conducted using the default inference configurations.

\textbf{Proteina.} We used the code and checkpoints from the official Proteina repository\footnote{\url{https://github.com/NVIDIA-Digital-Bio/proteina}}. Similar to La-Proteina, we adopted a length-dependent checkpoint strategy: \texttt{proteina\_v1.1\_DFS\_200M\_tir.ckpt} was used for lengths 100--500, and \texttt{proteina\_v1.6\_DFS\_200M\_notri\_long\_chain\_generation.ckpt} was used for lengths 600--1000. Default sampling parameters were applied in all cases.

\textbf{RFDiffusion.} We employed the official codebase available at the RFDiffusion repository\footnote{\url{https://github.com/RosettaCommons/RFdiffusion}}. All unconditional generation tasks were performed using the standard default sampling configurations provided by the authors.

\subsection{Binder Design Benchmark}
For the binder design task, all methods utilize identical target structures (in mmCIF format) and are conditioned on the exact same set of hotspot residues. For each target, we generate an equal number of candidate proteins with consistent lengths across all models. To ensure a fair comparison of generative capabilities, we disable any built-in auxiliary modules—such as pre-filtering, ranking, or early-stopping strategies—thereby restricting the evaluation strictly to the core sampling function.

\textbf{BindCraft.} We utilized the official implementation from the BindCraft repository\footnote{\url{https://github.com/martinpacesa/BindCraft}}. For binder generation, we adopted the \texttt{default\_4stage\_multimer} configuration under the advanced setting. To ensure a fair evaluation of the generative capability without post-hoc selection bias, we disabled both the filtering stage and the early stopping strategy. This ensured that every sampling attempt was completed and all raw outputs were retained, regardless of their internal quality metrics.

\textbf{BoltzGen.} We performed evaluations using the code provided in the BoltzGen repository\footnote{\url{https://github.com/HannesStark/boltzgen}}, specifically utilizing commit SHA \texttt{58c1eed}. We employed the \texttt{protein-anything} protocol for inference. To bypass internal ranking and filtering mechanisms, we set the inference budget equal to the number of desired designs, ensuring a direct one-to-one output.

\textbf{RFDiffusion.} For binder design tasks, we set both \texttt{noise\_scale\_ca} and \texttt{noise\_scale\_frame} to 0 during sampling. The authors suggest that this deterministic setting yields superior performance for specific binder design scenarios.

\textbf{RFDiffusion3.} We utilized the implementation available in the RosettaCommons Foundry repository\footnote{\url{https://github.com/RosettaCommons/foundry}}, specifically commit SHA \texttt{7f6656e}. We set \texttt{align\_trajectory\_structures=False} to prevent potential runtime errors during trajectory processing; all other configurations were maintained at default values.

\textbf{PXDesign.} We utilized the implementation in the PXDesign repository\footnote{\url{https://github.com/bytedance/PXDesign}}, specifically commit SHA \texttt{ec6615c}. We perform binder design for each target of a given hotspot using \texttt{Generation Only} mode.

\section{In Vitro Experimental Protocols}
\label{app:invitro_protocols}

\subsection{Protein Production}
\label{app:protein_production}

\paragraph{PD-L1.}
The extracellular domain of PD-L1 (residues 18--239) was cloned into pcDNA3.1 and expressed as a secreted protein in Expi293F cells via transient transfection with polyethylenimine (PEI). Transfection was performed when suspension cultures grown at 37~$^\circ$C reached a density of $3 \times 10^6$~cells/mL. Sodium butyrate (10~mM) was supplemented 9~hours post-transfection to enhance expression. Culture supernatants were harvested 5~days post-transfection. The pH was adjusted by adding 20~mM bis-tris propane (BTP) before loading onto Ni Sepharose Excel resin. The resin was washed with 10~column volumes of wash buffer (20~mM Na\textsubscript{2}HPO\textsubscript{4}/NaH\textsubscript{2}PO\textsubscript{4}, 200~mM NaCl, 20~mM imidazole, pH~7.4) and eluted with elution buffer (20~mM Na\textsubscript{2}HPO\textsubscript{4}/NaH\textsubscript{2}PO\textsubscript{4}, 200~mM NaCl, 300~mM imidazole, pH~7.4). The eluate was dialyzed and incubated with TEV protease for tag cleavage. The digested sample was passed over a Ni resin column; the flow-through was collected, concentrated, and further purified by gel-filtration chromatography on a Superdex~75 column equilibrated with SEC buffer (10~mM Na\textsubscript{2}HPO\textsubscript{4}, 1.8~mM KH\textsubscript{2}PO\textsubscript{4}, 137~mM NaCl, 2.7~mM KCl, pH~7.4). Peak fractions were pooled and concentrated to the desired concentration for downstream assays.

\paragraph{SARS-CoV-2 RBD.}
The SARS-CoV-2 receptor-binding domain (SC2RBD) was obtained commercially from AcroBiosystems (Cat.~\#~SPD-C82E9).

\paragraph{Minibinder Expression and Purification.}
Minibinder sequences were cloned with an N-terminal His-SUMO tag into a custom expression vector and transformed into BL21-CodonPlus~(DE3)-RIPL \textit{E.~coli}. Cultures were grown in Terrific Broth at 37~$^\circ$C until OD\textsubscript{600} reached 0.75, then induced with 1~mM isopropyl-$\beta$-D-thiogalactopyranoside (IPTG) for 4~hours at 37~$^\circ$C. Cells were lysed by high-pressure homogenization in lysis buffer (50~mM HEPES pH~8.0, 100~mM NaCl, 1~mM TCEP, 1~mM EDTA, supplemented with cOmplete EDTA-free protease inhibitor at 1~tablet per 100~mL). Proteins were purified by Ni-NTA chromatography, with the column washed using lysis buffer supplemented with 30~mM imidazole and eluted with 250~mM imidazole. The eluate was subjected to size-exclusion chromatography on a Superdex~200 10/300~GL column in 50~mM HEPES pH~8.0, 100~mM NaCl, 1~mM TCEP, 1~mM EDTA. Peak fractions were collected and stored for subsequent binding assays.

\subsection{Surface Plasmon Resonance (SPR)}
\label{app:spr}

SPR experiments were performed on a Biacore~8K+ instrument (GE Healthcare) at 25~$^\circ$C. Target proteins were immobilized on a Series~S Sensor Chip~SA (streptavidin, GE Healthcare) via biotin capture. The assay buffer consisted of 50~mM HEPES, 100~mM NaCl, 1~mM TCEP, and 1~mM EDTA at pH~8.0. The chip surface was activated with 1~M NaCl/50~mM NaOH (contact time: 60~s, flow rate: 30~$\mu$L/min), after which the biotinylated target protein was prepared at 1~$\mu$g/mL in assay buffer and coupled to the streptavidin surface to a density of approximately 100~Response Units (RU).

Minibinder stock solutions were maintained in the same assay buffer. For primary screening, each minibinder was tested at $10\times$ and $50\times$ dilutions of the stock concentration, along with a zero-concentration reference for background subtraction. Candidates were selected for affinity testing based on curve shape and RU response.

For affinity determination, either single-cycle kinetics (contact time: 70~s, dissociation time: 1200~s, flow rate: 30~$\mu$L/min, 9-point 1:1 serial dilution) or multi-cycle kinetics (contact time: 70~s, dissociation time: 600~s, flow rate: 30~$\mu$L/min, 12-point 1:1 serial dilution) were employed based on the estimated affinity of each minibinder. Data were analyzed using Biacore Insight Evaluation Software (version 6.0.7.1750). Kinetic parameters were determined by fitting sensorgrams to the Langmuir 1:1 binding model, which assumes a single-site interaction between the analyte and the immobilized ligand.

%% file: tables/training_pipeline.tex
\begin{table}[t]
\centering
\caption{\textbf{Training Stages.} }
\label{tab:training_stages}
\begin{tabular}{l c c c}
\toprule
\textbf{Hyperparameter} & \textbf{\makecell{Stage 1 \\ (Initial)}} & \textbf{\makecell{Stage 2 \\ (Fine-tuning 1)}} & \textbf{\makecell{Stage 3 \\ (Fine-tuning 2)}} \\
\midrule
Crop Size & 384 & 768 & 768 \\
Batch Size & 128 & 64 & 64 \\
Motif Percentage & 0\% & 20\% & 20\% \\
Training Steps & 50K & 20K & 30K \\
\midrule
\multicolumn{4}{l}{\textit{Data Sampling Distribution}} \\
\hspace{1em} Strictly Filtered Monomer & 100\% & 20\% & 10\% \\
\hspace{1em} Expanded Monomer Data & -- & 80\% & 40\% \\
\hspace{1em} Multimer Data & -- & -- & 50\% \\
\bottomrule
\end{tabular}
\end{table}

%% file: tables/targets.tex
\begin{table}[t]
\centering
\caption{\textbf{Benchmark Dataset for Binder Design.} 
We evaluated performance on 10 diverse targets. The table lists the target details, including the PDB ID, the specific chains/residues used as the target region, the key hotspot residues defining the binding site, and the range of binder lengths sampled during inference.}
\label{tab:targets}
\small 
\begin{tabularx}{\textwidth}{l l >{\raggedright\arraybackslash}X >{\raggedright\arraybackslash}X c}
\toprule
\textbf{Target} & \textbf{PDB ID} & \textbf{Target Region} & \textbf{Hotspot Residues} & \textbf{Length Range} \\
\midrule
\textbf{BHRF1}      & \texttt{2wh6} & Chain A: 2-158 & A65, A74, A77, A82, A85, A93 & 80-120 \\
\textbf{SC2RBD}     & \texttt{6m0j} & Chain E: 333-526 & E485, E489, E494, E500, E505 & 80-120 \\
\textbf{IL-7RA}     & \texttt{3di3} & Chain B: 17-209 & B58, B80, B139 & 50-120 \\
\textbf{PD-L1}      & \texttt{5o45} & Chain A: 17-132 & A56, A115, A123 & 50-120 \\
\textbf{TrkA}       & \texttt{1www} & Chain X: 282-382 & X294, X296, X333 & 50-120 \\
\textbf{Insulin}    & \texttt{4zxb} & Chain E: 6-155 & E64, E88, E96 & 40-120 \\
\textbf{H1}         & \texttt{5vli} & Chain A: 1-50, 76-80, 107-111, 258-322; Chain B: 1-68, 80-170 & B21, B45, B52 & 40-120 \\
\textbf{VEGF-A}     & \texttt{1bj1} & Chain V: 14-107; Chain W: 14-107 & W81, W83, W91 & 50-140 \\
\textbf{IL-17A}     & \texttt{4hsa} & Chain A: 17-131; Chain B: 19-127 & A94, A116, B67 & 50-140 \\
\textbf{TNF$\alpha$}& \texttt{1tnf} & Chain A/B/C: 12-157 & A113, C73 & 50-120 \\
\bottomrule
\end{tabularx}
\end{table}